\documentclass[aps,twocolumn,nofootinbib]{revtex4}%
\usepackage{amsfonts}
\usepackage{dcolumn}
\usepackage{graphicx}
\usepackage{amsmath}
\usepackage{amssymb}%
\begin{document}
\title[Surface tension of water]{Capillary waves at the liquid-vapor interface and the surface tension of water models}
\author{Ahmed E. Ismail, Gary S. Grest, and Mark J. Stevens}
\affiliation{Sandia National Laboratories, Albuquerque, NM\ 87185}
\begin{abstract}
Capillary waves occurring at the liquid-vapor interface of water are studied
using molecular dynamics simulations. In addition, the surface tension,
determined thermodynamically from the difference in the normal and tangential
pressure at the liquid-vapor interface, is compared for a number of standard
three- and four-point water models. We study four three-point models (SPC/E,
TIP3P, TIP3P-CHARMM, and TIP3P-Ew) and two four-point models (TIP4P and
TIP4P-Ew). All of the models examined underestimate the surface tension; the
TIP4P-Ew model comes closest to reproducing the experimental data. The surface
tension can also be determined from the amplitude of capillary waves at the
liquid-vapor interface by varying the surface area of the interface. The
surface tensions determined from the amplitude of the logarithmic divergence
of the capillary interfacial width and from the traditional thermodynamic
method agree only if the density profile is fitted to an error function
instead of a hyperbolic tangent function.

\end{abstract}
\date{\today}
\maketitle

\section{Introduction}

The ability to derive accurate property predictions for the liquid-vapor
interface is a key test for an atomistic force field. Because of the frequent
occurrence of water in systems of chemical and biological interest,
interfacial property prediction is especially vital for force fields of water.
The most important of these properties is surface tension, an intensive
quantity that measures the differential surface work required to increase the
interfacial area. Accurate models of the surface tension of water are
essential for conducting large-scale simulations of the wetting and spreading
of water droplets at surfaces.

An interface between two distinct thermodynamic phases can be characterized by
a local gradient of an order parameter whose mean value changes between
phases, such as the boundary between a liquid and its own vapor below the
critical temperature $T_{c}$. For simple fluids, thermodynamic arguments
predict that the interfacial width $\Delta$ depends only on temperature and
the interaction energies within each phase and across the interface. However,
the presence of the interface breaks the translational invariance of the
system, inducing Goldstone fluctuations or \textquotedblleft capillary
waves\textquotedblright\ at the interface \cite{Buff65,Rowlinson82}. Density
functional studies suggest that the surface tension predicted by
capillary-wave theory exhibits a minimum as a function of the wavelength
\cite{Mecke99}. Previous studies of capillary waves involving water have
tended to focus on liquid-liquid interfaces or on model fluids
\cite{Senapati01,Rivera03}, and have generally examined relatively small
systems of less than $10,000$ molecules; the present study represents the
first study of capillary-wave behavior at the liquid-vapor interface of water.

For two-dimensional interfaces, these noncritical fluctuations give rise to a
logarithmic increase in the interfacial width $\Delta$ with increasing
$L_{\parallel}$, the length of the interface. Most previous simulations
\cite{Nijmeijer88,Adams91} of the liquid-vapor interface in three dimensions
did not investigate the dependence of $\Delta$ on the size of the interface.
The purpose of this paper is to present atomistic molecular dynamics (MD)
simulations of the liquid-vapor interface of water. In particular, we obtain
the surface tension $\gamma$ in two different ways: from the difference in
pressure parallel $p_{\parallel}$ and perpendicular $p_{\perp}$ to the
interface ($\gamma_{p}$), and from from the dependence of $\Delta$ on
$L_{\parallel}$ ($\gamma_{w}$). We confirm the previous result that
$\gamma_{w}$ depends on the functional form chosen to fit the order parameter
(density profile) through the interface \cite{Lacasse98}. In particular,
fitting the order parameter to an error function gives results for $\gamma
_{w}$ which are in strong agreement with $\gamma_{p}$. However, fitting our
data to a hyperbolic tangent function, a functional form derived from
mean-field arguments \cite{Rowlinson82}, gives results for $\gamma_{w}$ which
are systematically smaller than $\gamma_{p}$ and further away from
experimental results.

There are currently a large number of different atomic models for water.
Guillot provides an extensive list of models developed through 2001
\cite{Guillot02}; several additional models have been introduced since then
\cite{Rivera02,Price04,Horn04,Jorgensen05}. The simplest of the commonly-used
atomic models, the SPC model \cite{Berendsen81}, is a rigid three-point model
with fixed charges; the most complex model, the POL5\ model \cite{Stern01}, is
a polarizable five-point model. Most of the commonly used models are three-
and four-point models. In three-point models, such as SPC/E \cite{Berendsen87}
and TIP3P \cite{Jorgensen83}, the electric charges are assigned directly to
the hydrogen and oxygen atoms; four-point models, such as the TIP4P
\cite{Jorgensen83} and Watanabe-Klein \cite{Watanabe89} models, locate the
negative charge at a massless point a fixed distance away from the oxygen
atom. Five-point models, such as TIP5P \cite{Mahoney00}, and the early
Bernal-Fowler \cite{Bernal33} and ST2 \cite{Stillinger74} models, represent
the negative charge of the oxygen using a pair of massless charges to capture
the quadrupolar behavior of water. Polarizable models, including the SPC/FQ
and TIP4P/FQ models \cite{Rick94}, allow the magnitude of the point charges to
be treated as variables which can fluctuate according to the local environment.

The proliferation of models has been motivated largely by the need to
reproduce various physical and thermodynamic properties, such as the bulk
density, the oxygen-oxygen radial distribution function, the heat of
vaporization, and the diffusion coefficient. However, some models, such as the
recent TIP3P-Ew \cite{Price04} and TIP4P-Ew models \cite{Horn04}, are
reparameterizations of existing models designed to account for changes in the
treatment of long-ranged electrostatic interactions.

Most of the available water models adequately represent at least some of the
thermodynamic properties of water; for a comprehensive review, see Jorgensen
\emph{et al.} \cite{Jorgensen05}. Kuo \emph{et al.} have shown that the
changes introduced between, for example, the TIP4P and TIP4P/FQ models have
little influence on properties such as the bulk liquid density or the mean
distance between oxygen atoms either in bulk or at the interface \cite{Kuo06}.
However, the simulation behavior of models with nearly identical parameters
can be markedly different: Mark and Nilsson have noted significant variation
in physical and thermodynamic properties such as the self-diffusion constant
and the radial distribution function of various three-point water models
\cite{Mark01,Mark02}. Less is known about how well the various water models
describe the liquid-vapor interface and the surface tension $\gamma$. Our own
work, however, suggests that even models with very similar density and
distribution profiles can have quite different predictions for surface tension.

Experimental studies demonstrate that the surface tension of water decreases
with a slight quadratic dependence on temperature in the range $273%
%TCIMACRO{\unit{K}}%
%BeginExpansion
\operatorname{K}%
%EndExpansion
<T<373%
%TCIMACRO{\unit{K}}%
%BeginExpansion
\operatorname{K}%
%EndExpansion
$ \cite{Gittens69,Cini72,Jasper72,Johansson72,Thakur75}. Surface tension
results for higher temperatures have not been reported in the literature; we
extrapolate the reported experimental data to higher temperatures. There have
been a few studies for various three-point models
\cite{Alejandre95,Feller96,Taylor96,Dang97,Zakharov97,Zakharov98,Rivera02,Shi06}
which show that while most water models reproduce the observed decrease in
surface tension as temperature increases, they tend to underestimate $\gamma$
by amounts between $25$ and $50$ percent. Only Alejandre \emph{et al.}, Huang
\emph{et al.}, and Shi \emph{et al. }\cite{Alejandre95,Huang01,Shi06} report
adequate agreement with experimental data. However, as we show below, the
apparent agreement of both Alejandre \emph{et al.} \cite{Alejandre95} and Shi
\emph{et al}. \cite{Shi06} is the result of inadequate simulation time.
Alejandre \emph{et al.} also employ a reciprocal-space mesh that is too
coarse, while Huang \emph{et al.} report values only for the SPC and
SPC/E\ models at $298%
%TCIMACRO{\unit{K}}%
%BeginExpansion
\operatorname{K}%
%EndExpansion
$ \cite{Huang01}.

Our primary goal is to study capillary waves at the liquid-vapor interface of
water, and to distinguish between various functional representations for the
density profile near the interface. Additionally, we first determine the
surface tension as a function of temperature for six commonly used three- and
four-point models of water, in part to establish a basis for comparison with
the capillary-wave simulations.

In Section II, we provide a brief overview of methods for computing the
surface tension from molecular simulation data, of the various water models
examined in this study, and of the simulation methods employed. Section III
presents our findings on the temperature dependence of the surface tension, as
well as the effects of the tail correction, interaction cutoffs, and
reciprocal-space mesh refinement. We discuss the results obtained from the
analysis of capillary waves at the liquid-vapor interface in Section
IV\ before offering our conclusions in\ Section\ V.

\section{\label{mm}Models and methodology}%

%TCIMACRO{\TeXButton{Table1}{\begingroup\newcolumntype{d}[0]{D{.}{.}{5}}
%\begin{table*}[tb]
%\caption{Parameters for commonly used three- and four-point models of water}
%\label{table1}
%\begin{tabular*}{\textwidth}{@{\extracolsep{\fill}}p{1.0in}cccccc}
%\hline\hline Parameter & {\textrm SPC/E} & {\textrm TIP3P} & {\textrm TIP3P-C}
%&
%{\textrm TIP3P-Ew} & {\textrm TIP4P} & {\textrm TIP4P-Ew} \\
%\end{tabular*}
%\begin{tabular*}{\textwidth}{@{\extracolsep{\fill}}p{1.0in}dddddd}
%\hline$q_H$ & 0.410 & 0.417 & 0.417 & 0.415 & 0.520 & 0.5242 \\
%$q_O$ & -0.820 & -0.834 & -0.834 & -0.830 & & \\
%$q_M$ & & & & & -1.040 & -1.0484 \\
%$\theta_{HOH}$, deg & 109.47 & 104.52 & 104.52 & 104.52 & 104.52 & 104.52\\
%$l_{OH}$, \AA & 1.0 & 0.9572 & 0.9572 & 0.9572 & 0.9572 & 0.9572 \\
%$l_{OM}$, \AA & & & & & 0.1500 & 0.1250 \\
%$\epsilon_{OO}%
%$, kcal/mol & 0.1553 & 0.1521 & 0.1521 & 0.102 & 0.1550 & 0.16275 \\
%$\sigma_{OO}$, \AA & 3.166 & 3.1506 & 3.1507  & 3.188 & 3.1536 & 3.16435 \\
%$\epsilon_{OH}$, kcal/mol & & & 0.0836 & & & \\
%$\sigma_{OH}$, \AA & & & 1.7753 & & & \\
%$\epsilon_{HH}$, kcal/mol & & & 0.0460 & & & \\
%$\sigma_{HH}$, \AA & & & 0.4000 & & & \\
%\hline\hline\end{tabular*}
%\end{table*}
%\endgroup}}%
%BeginExpansion
\begingroup\newcolumntype{d}[0]{D{.}{.}{5}}
\begin{table*}[tb]
\caption{Parameters for commonly used three- and four-point models of water}
\label{table1}
\begin{tabular*}{\textwidth}{@{\extracolsep{\fill}}p{1.0in}cccccc}
\hline\hline Parameter & {\textrm SPC/E} & {\textrm TIP3P} & {\textrm TIP3P-C}
&
{\textrm TIP3P-Ew} & {\textrm TIP4P} & {\textrm TIP4P-Ew} \\
\end{tabular*}
\begin{tabular*}{\textwidth}{@{\extracolsep{\fill}}p{1.0in}dddddd}
\hline$q_H$ & 0.410 & 0.417 & 0.417 & 0.415 & 0.520 & 0.5242 \\
$q_O$ & -0.820 & -0.834 & -0.834 & -0.830 & & \\
$q_M$ & & & & & -1.040 & -1.0484 \\
$\theta_{HOH}$, deg & 109.47 & 104.52 & 104.52 & 104.52 & 104.52 & 104.52\\
$l_{OH}$, \AA & 1.0 & 0.9572 & 0.9572 & 0.9572 & 0.9572 & 0.9572 \\
$l_{OM}$, \AA & & & & & 0.1500 & 0.1250 \\
$\epsilon_{OO}%
$, kcal/mol & 0.1553 & 0.1521 & 0.1521 & 0.102 & 0.1550 & 0.16275 \\
$\sigma_{OO}$, \AA & 3.166 & 3.1506 & 3.1507  & 3.188 & 3.1536 & 3.16435 \\
$\epsilon_{OH}$, kcal/mol & & & 0.0836 & & & \\
$\sigma_{OH}$, \AA & & & 1.7753 & & & \\
$\epsilon_{HH}$, kcal/mol & & & 0.0460 & & & \\
$\sigma_{HH}$, \AA & & & 0.4000 & & & \\
\hline\hline\end{tabular*}
\end{table*}
\endgroup
%EndExpansion

\subsection{Surface tension}

\subsubsection{Thermodynamic method}

\begin{figure}
[tb]
\begin{center}
\includegraphics[
trim=0.000000in 0.000000in -0.001146in 0.000000in,
height=1.4795in,
width=3.4255in
]%
{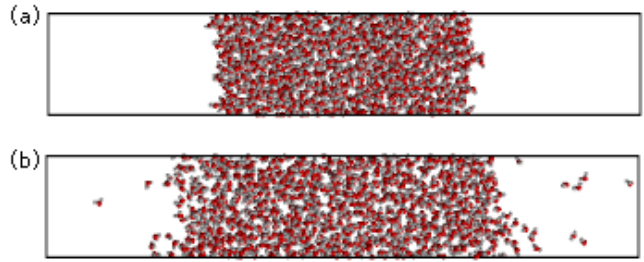}%
\caption{(Color online) Sample simulation cell, showing equilibrated
configurations of 1000 SPC/E water molecules at (a)\ $300\operatorname{K}$ and
(b) $500\operatorname{K}$. The dimensions of the cell are $L_{x}%
=L_{y}=L_{\parallel}=2.3\operatorname{nm}$ and $L_{z}=L_{\perp}%
=13.5\operatorname{nm}$.}%
\label{fig5}%
\end{center}
\end{figure}

There are two primary methods used to compute the surface tension using
molecular simulations. The first approach, developed by Tolman \cite{Tolman48}
and refined by Kirkwood and Buff \cite{Kirkwood49}, computes the surface
tension as an integral of the difference between the normal and tangential
pressures $p_{\perp}\left(  z\right)  $ and $p_{\parallel}\left(  z\right)  $:%
\begin{equation}
\gamma_{p}=\frac{1}{2}\int_{-\infty}^{\infty}\left(  p_{\perp}\left(
z\right)  -p_{\parallel}\left(  z\right)  \right)  ~dz, \label{eqn13}%
\end{equation}
where, in our geometry (see Figure \ref{fig5}),
\begin{align*}
p_{\perp}\left(  z\right)   &  =p_{z}\left(  z\right)  ,\\
p_{\parallel}\left(  z\right)   &  =\left(  p_{x}\left(  z\right)
+p_{y}\left(  z\right)  \right)  /2.
\end{align*}
The dominant contributions to the integral in Eq. (\ref{eqn13}) occur near the
interface; in the bulk away from the interface, $p_{\perp}=p_{\parallel}$ and
the integrand vanishes. For the specific case shown in Figure \ref{fig5},
where the interface separates a bulk liquid from its corresponding vapor
phase, the integral in Eq. (\ref{eqn13}) can be replaced with an ensemble
average of the difference between the normal and tangential pressures,%
\begin{equation}
\gamma_{p}=\frac{L_{z}}{2}\left\langle p_{\perp}-p_{\parallel}\right\rangle
=\frac{L_{z}}{2}\left[  \left\langle p_{z}\right\rangle -\frac{\left\langle
p_{x}\right\rangle +\left\langle p_{y}\right\rangle }{2}\right]  .
\label{eqn14}%
\end{equation}
The outer factor of $1/2$ in Eq. (\ref{eqn14})\ accounts for the presence of
two liquid-vapor interfaces.%

%TCIMACRO{\FRAME{ftbFU}{3.4255in}{1.4795in}{0pt}{\Qcb{(Color online) Sample
%simulation cell, showing equilibrated configurations of 1000 SPC/E water
%molecules at (a)\ $300\unit{K}$ and (b) $500\unit{K}$. The dimensions of the
%cell are $L_{x}=L_{y}=L_{\parallel}=2.3\unit{nm}$ and $L_{z}=L_{\perp
%}=13.5\unit{nm}$.}}{\Qlb{fig5}}{simcell300_500.eps}%
%width 3.4255in;  height 1.4795in;  depth 0pt;  original-width 3.8193in;
%original-height 1.632in;  cropleft "0";  croptop "1";  cropright "1.0003";
%cropbottom "0";  filename 'simcell300_500.eps';file-properties "XNPEU";}}}%
%BeginExpansion
%EndExpansion

In most numerical simulations, the interatomic and electrostatic interactions
are only applied within a cutoff range $r_{c}$. The introduction of the cutoff
in the interparticle potential reduces the surface tension in much the same
way that the introduction of a cutoff reduces the bulk pressure at constant
density. Thus, the simulation result $\gamma_{p}$ will underestimate the total
surface tension; a better estimate of the total surface tension can be
obtained from%
\begin{equation}
\gamma=\gamma_{p}+\gamma_{tail}, \label{eqn23}%
\end{equation}
where $\gamma_{tail}$, the tail correction for $\gamma_{p}$, can be determined
from \cite{Chapela77,Blokhuis95}%
\begin{multline}
\gamma_{tail}=\frac{\pi}{2}\int_{-\infty}^{\infty}\int_{-1}^{1}\int_{r_{c}%
}^{\infty}r^{3}U^{\prime}\left(  r\right)  g\left(  r\right)  \left(
1-3s^{2}\right)  \times\label{eqn15}\\
\left(  \rho\left(  z\right)  \rho\left(  z-sr\right)  -\left(  \rho
_{G}\left(  z\right)  \right)  ^{2}\right)  ~dr~ds~dz,
\end{multline}
where $U\left(  r\right)  $ is the pairwise potential, $g\left(  r\right)  $
is the radial distribution function, $\rho\left(  z\right)  $ is the observed
interfacial profile, and $\rho_{G}\left(  z\right)  $ is a Gibbs dividing
surface:%
\begin{equation}
\rho_{G}\left(  z\right)  =\rho_{c}+\frac{\Delta\rho}{2}\operatorname*{sgn}%
\left(  z\right)  . \label{eqn16}%
\end{equation}
Although the use of the tail correction in Eq. (\ref{eqn23}) improves the
estimate of the surface tension, its use is restricted to systems in which the
two phases contain the same components; for composite systems, such as water
at the surface of a solid, only $\gamma_{p}$ should be used.

We assume in Eqs. (\ref{eqn15})\ and (\ref{eqn16}) that the liquid-vapor
interface is centered at $z=0$. In Eq. (\ref{eqn16}), $\rho_{c}=\left(
\rho_{l}+\rho_{v}\right)  /2$ is the average density of the two phases, and
$\Delta\rho=\rho_{l}-\rho_{v}$ is the difference between the average densities
of the two phases. Thus, $\rho_{G}\left(  z\right)  =\rho_{v}$ for $z<0$ and
$\rho_{l}$ for $z>0$. There are multiple possible choices for determining the
observed density profile $\rho\left(  z\right)  $. Although it is possible to
use the profile calculated from the simulation directly, both the
tail-correction and capillary-wave calculations are simplified by fitting the
profile to a function. In the present study, the density profile is fitted to
both an error function and a hyperbolic tangent function, as discussed in the
following section.

\subsubsection{\label{cwm}Capillary-wave method}

The thermodynamic approach for computing surface tension assumes a sharp
liquid-vapor interface when in reality it is quite rough. The roughness of the
interface increases at high temperatures, as seen in Figure \ref{fig5}. A
second method for computing the surface tension assumes that the observed
magnitude of the fluctuations is derived from two sources:\ an intrinsic
contribution plus a logarithmic term that represents broadening of the
interface as a result of the capillary waves
\cite{Buff65,Huang69,Beysens87,Semenov94,Lacasse98,Sides99}.

If the contributions from capillary waves can be decoupled from density
fluctuations, then the surface tension can be computed by determining the
interfacial profiles for a number of different system sizes. The relationship
between the observed interfacial width $\Delta$ and the intrinsic interfacial
width $\Delta_{0}$ is given by%
\begin{equation}
\Delta^{2}=\Delta_{0}^{2}+\frac{k_{B}T}{2\pi\gamma_{w}}\ln\left(
\frac{L_{\parallel}}{B_{0}}\right)  , \label{eqn17}%
\end{equation}
where $L_{\parallel}=L_{x}=L_{y}$ is the length of the interface, and $B_{0}$
is a characteristic length scale related to the short-wavelength cutoff in the
interfacial behavior. It is unnecessary to determine $B_{0}$ before computing
the surface tension $\gamma_{w}$.

Computation of $\gamma_{w}$ requires the scaled density profile,
\begin{equation}
\Psi\left(  z\right)  =\frac{2}{\rho_{L}-\rho_{V}}\left(  \rho\left(
z\right)  -\frac{\rho_{L}+\rho_{V}}{2}\right)  , \label{eqn21}%
\end{equation}
in the $z$-direction. Given $\Psi\left(  z\right)  $, the variance in the
derivative of the profile $f\left(  z\right)  =\Psi^{\prime}\left(  z\right)
$ can be computed in either real or reciprocal (Fourier) space:%
\begin{equation}
\Delta^{2}=\frac{\int_{-\infty}^{\infty}z^{2}f\left(  z\right)  ~dz}%
{\int_{-\infty}^{\infty}f\left(  z\right)  ~dz}=-\frac{1}{\tilde{f}\left(
0\right)  }\left[  \frac{d^{2}\tilde{f}\left(  q\right)  }{dq^{2}}\right]
_{q=0}, \label{eqn18}%
\end{equation}
where $\tilde{f}\left(  q\right)  $ is the Fourier transform of $f\left(
z\right)  $. The simple form of Eq. (\ref{eqn18}) suggests that fitting the
profile $\Psi\left(  z\right)  $ to a functional form will be both more
convenient and lead to more accurate results than using the raw profile data.
Several different functional forms for $\Psi\left(  z\right)  $ have been
proposed in the literature. Relying on mean-field arguments, most previous
theoretical and computational studies of surface tension have fitted the
profile to a hyperbolic tangent function
\cite{Chapela77,Blokhuis95,Alejandre95,Kuo06},
\begin{equation}
\Psi_{t}\left(  z\right)  =\tanh\left(  \frac{2z}{w_{t}}\right)  ,
\label{eqn19}%
\end{equation}
while Huang and Webb \cite{Huang69} and Beysens and Robert \cite{Beysens87}
propose the use of an error function,%
\begin{equation}
\Psi_{e}\left(  z\right)  =\operatorname{erf}\left(  \frac{\sqrt{\pi}z}{w_{e}%
}\right)  . \label{eqn20}%
\end{equation}
If the density profile $\Psi\left(  z\right)  $ is fitted to a hyperbolic
tangent function Eq. (\ref{eqn19}), then from Eq. (\ref{eqn18}) we find that
\cite{Lacasse98}
\[
\Delta_{t}^{2}=\pi^{2}w_{t}^{2}/48,
\]
while for an error function Eq. (\ref{eqn20}), the interfacial width
$\Delta_{e}^{2}$ is given by%
\[
\Delta_{e}^{2}=w_{e}^{2}/2\pi.
\]
We will show that there is a significant discrepancy between the surface
tensions obtained from the hyperbolic tangent profile, Eq. (\ref{eqn19}), and
the error function profile, Eq. (\ref{eqn20}), with the error function giving
results in closer agreement with Eq. (\ref{eqn14}).

\subsection{Water models}

We consider four different three-point models: the SPC/E model; the original
TIP3P model; the modification of the TIP3P model \cite{Durell94} implemented
in CHARMM (hereafter referred to as TIP3P-C); and the TIP3P-Ew model
\cite{Price04}, a recent reparameterization incorporating the effects of Ewald
summation. The parameters for the different water models are summarized in
Table \ref{table1}.

The basic structure of the different models are similar. The common features
of all models include a specified oxygen-hydrogen bond length $l_{OH}$ and
hydrogen-oxygen-hydrogen bond angle $\theta_{HOH}$, a charge on each hydrogen
atom, and a Lennard-Jones 12-6 potential describing the interaction between
the oxygen atoms,%
\begin{equation}
U_{LJ}\left(  r_{OO}\right)  =\left\{
\begin{array}
[c]{cc}%
4\varepsilon_{OO}\left[  \left(  \frac{\sigma_{OO}}{r_{OO}}\right)
^{12}-\left(  \frac{\sigma_{OO}}{r_{OO}}\right)  ^{6}\right]  , & r_{OO}\leq
r_{c}\\
0, & r_{OO}>r_{c}%
\end{array}
\right.  , \label{eqn11}%
\end{equation}
where $\varepsilon_{OO}$ and $\sigma_{OO}$ are the model-dependent well depth
and equilibrium O-O distance, and $r_{OO}$ is the distance between oxygen
atoms. The TIP3P-C model incorporates hydrogen-hydrogen and hydrogen-oxygen
Lennard-Jones interactions as well.

In addition to the Lennard-Jones interaction, there are electrostatic
interactions between the charge sites:%
\begin{equation}
U_{es}\left(  r_{ij}\right)  =\sum_{i=1}^{N}\sum_{j=1}^{N}\frac{q_{i}q_{j}%
}{4\pi\varepsilon_{0}r_{ij}}, \label{eqn12}%
\end{equation}
where $q_{\alpha}$ is the charge on atom $\alpha$, and $r_{ij}$ is the
distance between atoms $i$ and $j$ in the simulation box. Previous studies
have shown that significant variations in the values obtained for surface
tension can occur depending upon how the sum in Eq. (\ref{eqn12}) is performed
\cite{Feller96}. Except in Section \ref{cutoff}, Ewald summations were used
throughout our simulations.

We also consider a pair of four-point water models: the TIP4P model
\cite{Jorgensen83}, and the TIP4P-Ew model \cite{Horn04}, a recent
reparameterization designed to account for the presence of long-range
interactions. The four-point models introduce a bare charge at a new site,
designated $M$, located on the bisector of the HOH bond angle; the charge is
of strength $q_{M}$. The forces acting on the massless site are distributed to
the O and H atoms in the same molecule \cite{Feenstra99}:%
\begin{align*}
\mathbf{F}_{ij,O}  &  =\left(  1-2a\right)  \mathbf{F}_{ij,M},\\
\mathbf{F}_{ij,H}  &  =a\mathbf{F}_{ij,M},
\end{align*}
where $a=l_{OM}/\left(  l_{OH}\cos\left(  \theta_{HOH}/2\right)  \right)  $
and $\mathbf{F}_{ij,M}$ is the force acting on the massless site associated
with oxygen $i$ due to atom $j$ \cite{Spoel05} \footnote{In the case of
four-point water models, while the contribution to the virial from the
Lennard-Jones interactions can be computed using the more computationally
efficient form $\sum_{i}\mathbf{r}_{i}\cdot\mathbf{f}_{i}$, the contribution
to the virial from short-range electrostatic forces is computed using the
pairwise form $\sum_{i>j}\mathbf{r}_{ij}\cdot\mathbf{f}_{ij}$.}. For the TIP4P
model, the charge is located $l_{OM}=0.15%
%TCIMACRO{\unit{\mathring{A}}}%
%BeginExpansion
\operatorname{\mathring{A}}%
%EndExpansion
$ away from the oxygen atom. The TIP4P-Ew model changes the values of $l_{OM}%
$, the charge $q_{M}$, as well as the separation $\sigma_{OO}$ and well-depth
$\varepsilon_{OO}$ of the Lennard-Jones interaction.

\subsection{Simulation method}

\subsubsection{Thermodynamic method}

To determine the surface tension of the various three-point water models, 1000
molecules were placed into a periodic, rectangular box of dimensions
$L_{x}=L_{y}=L_{\parallel}=2.3%
%TCIMACRO{\unit{nm}}%
%BeginExpansion
\operatorname{nm}%
%EndExpansion
$ and $L_{z}=L_{\perp}=13.5%
%TCIMACRO{\unit{nm}}%
%BeginExpansion
\operatorname{nm}%
%EndExpansion
$. The increased system size in the $z$-direction minimizes the interactions
of water molecules in the liquid phase with their $z$-periodic images through
the long-range Coulombic interactions in Eq. (\ref{eqn12}). Similarly, 1000
molecules of the four-point models were simulated in a box with dimensions
$L_{x}=L_{y}=L_{\parallel}=2.7%
%TCIMACRO{\unit{nm}}%
%BeginExpansion
\operatorname{nm}%
%EndExpansion
$ and $L_{z}=L_{\perp}=12.0%
%TCIMACRO{\unit{nm}}%
%BeginExpansion
\operatorname{nm}%
%EndExpansion
$, each also containing $1000$ molecules. The initial configuration was
constructed by placing the water molecules at the center of a simple cubic
lattice with 7 molecules each in the $x$- and $y$-directions, and the
$z$-spacing chosen to create a density of $0.98%
%TCIMACRO{\unit{g}}%
%BeginExpansion
\operatorname{g}%
%EndExpansion%
%TCIMACRO{\unit{cm}}%
%BeginExpansion
\operatorname{cm}%
%EndExpansion
^{-3}$ for the three-point models, and $1.00%
%TCIMACRO{\unit{g}}%
%BeginExpansion
\operatorname{g}%
%EndExpansion%
%TCIMACRO{\unit{cm}}%
%BeginExpansion
\operatorname{cm}%
%EndExpansion
^{-3}$ for the four-point models. The same starting configuration was used for
all simulations of a given water model. At equilibrium, the thickness of the
slab in the $z$-direction varied between approximately $5.5%
%TCIMACRO{\unit{nm}}%
%BeginExpansion
\operatorname{nm}%
%EndExpansion
$ at $300%
%TCIMACRO{\unit{K}}%
%BeginExpansion
\operatorname{K}%
%EndExpansion
$ and $7.5%
%TCIMACRO{\unit{nm}}%
%BeginExpansion
\operatorname{nm}%
%EndExpansion
$ at $500%
%TCIMACRO{\unit{K}}%
%BeginExpansion
\operatorname{K}%
%EndExpansion
$.

For each of the six models examined, molecular dynamics (MD) simulations were
performed in the $NVT$ ensemble in $25$-degree increments between $300%
%TCIMACRO{\unit{K}}%
%BeginExpansion
\operatorname{K}%
%EndExpansion
$ and $500%
%TCIMACRO{\unit{K}}%
%BeginExpansion
\operatorname{K}%
%EndExpansion
$ using the LAMMPS\ simulation package \cite{Plimpton95}. The cutoff for the
Lennard-Jones potentials and the short-range cutoff for the electrostatic
potentials were set to $10%
%TCIMACRO{\unit{\mathring{A}}}%
%BeginExpansion
\operatorname{\mathring{A}}%
%EndExpansion
$, unless otherwise specified. The bond lengths and bond angles of the various
models were constrained using the SHAKE\ technique \cite{Ryckaert77}. The
equations of motion were integrated using the Verlet algorithm with velocity
rescaling to control the temperature. The difference in the surface tension
between simulations performed with velocity rescaling and those with a
Nos\'{e}-Hoover thermostat with a damping constant of $100%
%TCIMACRO{\unit{ps}}%
%BeginExpansion
\operatorname{ps}%
%EndExpansion
^{-1}$ was significantly less than the simulation uncertainty. Each simulation
was performed for a total of $2%
%TCIMACRO{\unit{ns}}%
%BeginExpansion
\operatorname{ns}%
%EndExpansion
$ with time step $\Delta t=1%
%TCIMACRO{\unit{fs}}%
%BeginExpansion
\operatorname{fs}%
%EndExpansion
.$ The system was allowed to equilibrate for $1%
%TCIMACRO{\unit{ns}}%
%BeginExpansion
\operatorname{ns}%
%EndExpansion
$; data from the second $1%
%TCIMACRO{\unit{ns}}%
%BeginExpansion
\operatorname{ns}%
%EndExpansion
$ were used to compute the surface tension.

The electrostatic interactions were calculated using the particle-particle
particle-mesh (PPPM) technique of Hockney and Eastwood \cite{Hockney88}. The
mesh spacing in this work was selected to ensure that the root-mean-squared
accuracy of the force calculation was within $10^{-4}$; the resulting grid was
of dimensions $12\times12\times48$. Most previous simulations were carried out
with a maximum of $h_{z}^{\max}=20$ cells in the $z$-direction
\cite{Alejandre95,Feller96}. In several of those studies, simulations were
carried out with $h_{z}^{\max}=10$ or less, and some did not include
long-range electrostatic interactions at all \cite{Taylor96}. We consider the
effects of mesh refinement on the surface tension in Section \ref{rsad}.

\subsubsection{\label{sm}Capillary-wave method}

Observation of capillary waves requires simulations with larger interfacial
surface areas than were used in the thermodynamic method above. Consequently,
we studied systems with $L_{x}=L_{y}=L_{\parallel}$ varying between $9.2%
%TCIMACRO{\unit{nm}}%
%BeginExpansion
\operatorname{nm}%
%EndExpansion
$ and $46.0%
%TCIMACRO{\unit{nm}}%
%BeginExpansion
\operatorname{nm}%
%EndExpansion
$. The resulting simulations used to compute the surface tension have surface
areas between $84.6%
%TCIMACRO{\unit{nm}}%
%BeginExpansion
\operatorname{nm}%
%EndExpansion
^{2}$ and $2116.0%
%TCIMACRO{\unit{nm}}%
%BeginExpansion
\operatorname{nm}%
%EndExpansion
^{2},$ and contained between $16,000$ and $400,000$ water molecules. To
construct the initial configuration, we take an equilibrated sample and
replicate it multiple times in the $x$- and $y$-directions. The SPC/E model
was used for this study, as it was the most computationally efficient of the
models studied.

\begin{figure}
[tb]
\begin{center}
\includegraphics[
height=2.2458in,
width=3.1985in
]%
{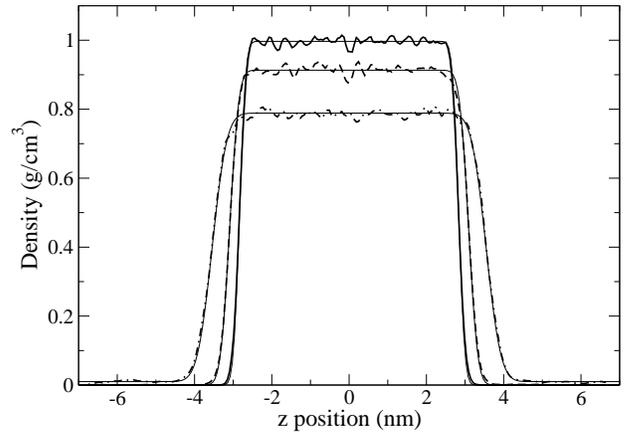}%
\caption{Density profiles for the SPC/E model of water at 300 K (thick solid
line), 400 K (thick dashed line), and 500 K (thick dashed-dotted line). Fits
to error functions are shown as thin solid lines.}%
\label{fig3}%
\end{center}
\end{figure}

To ensure that artifacts from the replication process were eliminated, the
simulation time varied between $1.0%
%TCIMACRO{\unit{ns}}%
%BeginExpansion
\operatorname{ns}%
%EndExpansion
$ and $6.0%
%TCIMACRO{\unit{ns}}%
%BeginExpansion
\operatorname{ns}%
%EndExpansion
$ as an increasing function of $L_{\parallel}$. Only the last $750%
%TCIMACRO{\unit{ps}}%
%BeginExpansion
\operatorname{ps}%
%EndExpansion
$ of the simulation were used for recording data; the preceding steps were
used for equilibration and discarded. We output the position of every atom
after every $5000$ timesteps, and then assigned each atom to one of 500 bins
depending on its location in the $z$-direction. To ensure that the interfacial
profile was not altered by drifts in the density profile, the profile was
shifted so that the center of the mass was located at $z=0$. Sample profiles
for the SPC/E model of water at several temperatures are shown in Figure
\ref{fig3}. After the average density profile $\rho\left(  z\right)  $ was
computed, the two halves of the profile, on either side of $z=0$, were
averaged together, rescaled to values between $-1$ and $1$ using Eq.
(\ref{eqn21}), and then fitted to hyperbolic tangent and error functions of
the form of Eq. (\ref{eqn19})\ and Eq. (\ref{eqn20}).%

%TCIMACRO{\FRAME{ftbFU}{3.1985in}{2.2458in}{0pt}{\Qcb{Density profiles for the
%SPC/E model of water at 300 K (thick solid line), 400 K (thick dashed line),
%and 500 K (thick dashed-dotted line). Fits to error functions are shown as
%thin solid lines.}}{\Qlb{fig3}}{bw_density.eps}%
%width 3.1985in;  height 2.2458in;  depth 0pt;  original-width 9.6734in;
%original-height 6.7739in;  cropleft "0";  croptop "1";  cropright "1";
%cropbottom "0";  filename 'bw_density.eps';file-properties "XNPEU";}}}%
%BeginExpansion
%EndExpansion
%

%TCIMACRO{\TeXButton{Table2}{\begin{table*}[tb]
%\caption
%{Surface tension$^a$ for three- and four-point water models, including tail correction}
%\label{table2}
%\begin{ruledtabular}
%\begin{tabular}{@{\extracolsep{\fill}}cccccccc}
%& \multicolumn{7}{c}{Surface Tension, $\gamma$ (mN/m)} \\
%\cline{2-8}
%Temperature (K) & \hspace{0.1in}SPC/E\hspace{0.1in} &
%\hspace{0.1in}TIP3P\hspace{0.1in} &
%\hspace{0.1in}TIP3P-Ew\hspace{0.1in} &
%\hspace{0.1in}TIP3P-C\hspace{0.1in} &
%\hspace{0.1in}TIP4P\hspace{0.1in} &
%\hspace{0.1in}TIP4P-Ew\hspace{0.1in} &
%Expt.$^b$ \\
%\hline300 & 55.4 & 51.1 & 47.4 & 48.8 & 53.6 & 61.2 & 71.7 \\
%325 & 47.9 & 45.9 & 43.6 & 47.5 & 50.9 & 55.6 & 67.6 \\
%350 & 47.0 & 42.8 & 39.2 & 45.7 & 45.7 & 52.7 & 63.2 \\
%375 & 44.6 & 37.8 & 37.7 & 39.9 & 41.4 & 48.1 & 58.4 \\
%400 & 37.6 & 35.5 & 34.3 & 36.9 & 35.9 & 43.5 & 53.3 \\
%425 & 32.0 & 31.5 & 28.9 & 31.9 & 31.2 & 38.6 & 47.9 \\
%450 & 30.6 & 27.3 & 25.9 & 28.2 & 25.7 & 34.5 & 42.1 \\
%475 & 26.8 & 24.7 & 23.2 & 23.2 & 19.1 & 29.3 & 36.0 \\
%500 & 23.2 & 17.0 & 16.9 & 18.1 & 15.2 & 24.8 & 29.5 \\
%\end{tabular}
%\end{ruledtabular}
%{\em Notes:}
%$^a$Uncertainty for all simulation results is between 2.4 and 3.0 mN/m.
%$^b$Experimental data taken from Refs. \cite{Gittens69,Cini72,Jasper72};
%data above 400 K is extrapolated from quadratic fit provided in
%Ref. \cite{Cini72}.
%\end{table*}}}%
%BeginExpansion
\begin{table*}[tb]
\caption
{Surface tension$^a$ for three- and four-point water models, including tail correction}
\label{table2}
\begin{ruledtabular}
\begin{tabular}{@{\extracolsep{\fill}}cccccccc}
& \multicolumn{7}{c}{Surface Tension, $\gamma$ (mN/m)} \\
\cline{2-8}
Temperature (K) & \hspace{0.1in}SPC/E\hspace{0.1in} &
\hspace{0.1in}TIP3P\hspace{0.1in} &
\hspace{0.1in}TIP3P-Ew\hspace{0.1in} &
\hspace{0.1in}TIP3P-C\hspace{0.1in} &
\hspace{0.1in}TIP4P\hspace{0.1in} &
\hspace{0.1in}TIP4P-Ew\hspace{0.1in} &
Expt.$^b$ \\
\hline300 & 55.4 & 51.1 & 47.4 & 48.8 & 53.6 & 61.2 & 71.7 \\
325 & 47.9 & 45.9 & 43.6 & 47.5 & 50.9 & 55.6 & 67.6 \\
350 & 47.0 & 42.8 & 39.2 & 45.7 & 45.7 & 52.7 & 63.2 \\
375 & 44.6 & 37.8 & 37.7 & 39.9 & 41.4 & 48.1 & 58.4 \\
400 & 37.6 & 35.5 & 34.3 & 36.9 & 35.9 & 43.5 & 53.3 \\
425 & 32.0 & 31.5 & 28.9 & 31.9 & 31.2 & 38.6 & 47.9 \\
450 & 30.6 & 27.3 & 25.9 & 28.2 & 25.7 & 34.5 & 42.1 \\
475 & 26.8 & 24.7 & 23.2 & 23.2 & 19.1 & 29.3 & 36.0 \\
500 & 23.2 & 17.0 & 16.9 & 18.1 & 15.2 & 24.8 & 29.5 \\
\end{tabular}
\end{ruledtabular}
{\em Notes:}
$^a$Uncertainty for all simulation results is between 2.4 and 3.0 mN/m.
$^b$Experimental data taken from Refs. \cite{Gittens69,Cini72,Jasper72};
data above 400 K is extrapolated from quadratic fit provided in
Ref. \cite{Cini72}.
\end{table*}%
%EndExpansion
%

%TCIMACRO{\TeXButton{Table3}{\begin{table*}[tbf]
%\caption
%{Interfacial properties of SPC/E water as a function of temperature, for $r_c = 10$ \AA and $L_{\parallel
%} = 2.3$ nm}
%\label{table3}
%\begin{ruledtabular}
%\begin{tabular}{ccccc}
%Temperature & Interface thickness  & Liquid density
%& Vapor density & Tail correction \\
%$T$, K & $w_e$,  \AA & $\rho_L$, g/cm$^3$ & $\rho_V$, g/cm$^3$ & $\gamma_{tail}
%$, mN/m \\
%\hline300 & 3.12 & 0.990 & 0.0005 & 5.5 \\
%325 & 3.37 & 0.977 & 0.0008 & 5.2 \\
%350 & 3.75 & 0.959 & 0.0005 & 5.0 \\
%375 & 4.22 & 0.941 & 0.0005 & 4.8 \\
%400 & 4.63 & 0.913 & 0.0015 & 4.5 \\
%425 & 5.23 & 0.886 & 0.0023 & 4.1 \\
%450 & 5.74 & 0.855 & 0.0049 & 3.8 \\
%475 & 6.00 & 0.818 & 0.0093 & 3.5 \\
%500 & 7.54 & 0.779 & 0.0199 & 3.0 \\
%\end{tabular}
%\end{ruledtabular}
%\end{table*}}}%
%BeginExpansion
\begin{table*}[tbf]
\caption
{Interfacial properties of SPC/E water as a function of temperature, for $r_c = 10$ \AA and $L_{\parallel
} = 2.3$ nm}
\label{table3}
\begin{ruledtabular}
\begin{tabular}{ccccc}
Temperature & Interface thickness  & Liquid density
& Vapor density & Tail correction \\
$T$, K & $w_e$,  \AA & $\rho_L$, g/cm$^3$ & $\rho_V$, g/cm$^3$ & $\gamma_{tail}
$, mN/m \\
\hline300 & 3.12 & 0.990 & 0.0005 & 5.5 \\
325 & 3.37 & 0.977 & 0.0008 & 5.2 \\
350 & 3.75 & 0.959 & 0.0005 & 5.0 \\
375 & 4.22 & 0.941 & 0.0005 & 4.8 \\
400 & 4.63 & 0.913 & 0.0015 & 4.5 \\
425 & 5.23 & 0.886 & 0.0023 & 4.1 \\
450 & 5.74 & 0.855 & 0.0049 & 3.8 \\
475 & 6.00 & 0.818 & 0.0093 & 3.5 \\
500 & 7.54 & 0.779 & 0.0199 & 3.0 \\
\end{tabular}
\end{ruledtabular}
\end{table*}%
%EndExpansion
%

%TCIMACRO{\TeXButton{Table4}{\begin{table}[b]
%\caption{Surface tensions $\gamma_p$ and $\gamma
%$ and liquid-phase density $\rho_L$ for the SPC/E model
%as a function of LJ cutoff $r_c$, with and without tail correction, at 300 K}
%\label{table4}
%\begin{ruledtabular}
%\begin{tabular}{@{\extracolsep{\fill}}cccc}
%$r_{\textit c}$ (\AA) & $\gamma_p$ (mN/m) & $\gamma$ (mN/m) & $\rho
%_L$ (g/cm$^3$) \\
%\hline10.0 & 46.3 & 51.8 & 0.990 \\
%12.0 & 51.2 & 55.0 & 0.992 \\
%14.0 & 47.9 & 50.6 & 0.994 \\
%16.0 & 49.7 & 51.8 & 0.996 \\
%18.0 & 49.9 & 51.5 & 0.996 \\
%20.0 & 52.8 & 54.1 & 0.995 \\
%\end{tabular}
%\end{ruledtabular}
%\end{table}}}%
%BeginExpansion
\begin{table}[b]
\caption{Surface tensions $\gamma_p$ and $\gamma
$ and liquid-phase density $\rho_L$ for the SPC/E model
as a function of LJ cutoff $r_c$, with and without tail correction, at 300 K}
\label{table4}
\begin{ruledtabular}
\begin{tabular}{@{\extracolsep{\fill}}cccc}
$r_{\textit c}$ (\AA) & $\gamma_p$ (mN/m) & $\gamma$ (mN/m) & $\rho
_L$ (g/cm$^3$) \\
\hline10.0 & 46.3 & 51.8 & 0.990 \\
12.0 & 51.2 & 55.0 & 0.992 \\
14.0 & 47.9 & 50.6 & 0.994 \\
16.0 & 49.7 & 51.8 & 0.996 \\
18.0 & 49.9 & 51.5 & 0.996 \\
20.0 & 52.8 & 54.1 & 0.995 \\
\end{tabular}
\end{ruledtabular}
\end{table}%
%EndExpansion

\section{Thermodynamic Surface Tension:\ Results and Discussion}

\subsection{\label{Tdep}Temperature dependence}

\begin{figure}
[tb]
\begin{center}
\includegraphics[
height=4.8837in,
width=3.4255in
]%
{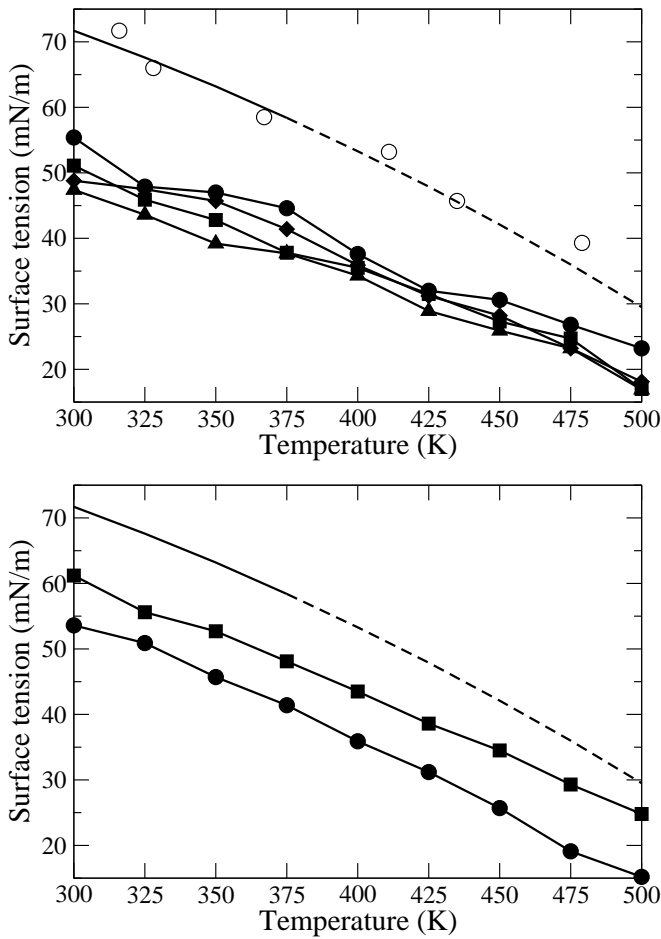}%
\caption{(Top figure) Surface tension of the three-point models of water as a
function of temperature: SPC/E (filled circles), TIP3P (squares), TIP3P-C
(triangles), and TIP3P-Ew (diamonds). Simulation data from Alejandre \emph{et
al.} \cite{Alejandre95} (open circles) are included for comparison. (Bottom
figure) Surface tension of the four-point models of water as a function of
temperature: TIP4P (circles) and TIP4P-Ew (squares). In both figures,
experimental data \cite{Gittens69,Cini72,Jasper72} (solid curve) and
extrapolation of quadratic fit to higher temperatures (dashed curve) are
included for comparison.}%
\label{fig3ab}%
\end{center}
\end{figure}

The surface tension $\gamma=\gamma_{p}+\gamma_{tail}$ of the various water
models, computed using Eq. (\ref{eqn14}), are collected in Table \ref{table2}.
Results for the three-point models are shown in Figure \ref{fig3ab}(a) as a
function of temperature. Using the method of Flyvbjerg and Petersen
\cite{Flyvbjerg89}, the uncertainty in the results was found to be between
$2.4$ and $3.0%
%TCIMACRO{\unit{mN}}%
%BeginExpansion
\operatorname{mN}%
%EndExpansion
/%
%TCIMACRO{\unit{m}}%
%BeginExpansion
\operatorname{m}%
%EndExpansion
$. Comparing the four three-point models, we find that SPC/E model is the
closest to the experimental data, with better agreement at higher
temperatures. For most temperatures, the various TIP3P models agree with the
SPC/E model within the uncertainty of the simulation. Although the three-point
models considered do not achieve agreement with experimental data, the overall
temperature dependence of $\gamma$ for the models is in good agreement with
the experimental data. For the three-point models, $\gamma$ is generally
between $15%
%TCIMACRO{\unit{mN}}%
%BeginExpansion
\operatorname{mN}%
%EndExpansion
/%
%TCIMACRO{\unit{m}}%
%BeginExpansion
\operatorname{m}%
%EndExpansion
$ and $20%
%TCIMACRO{\unit{mN}}%
%BeginExpansion
\operatorname{mN}%
%EndExpansion
/%
%TCIMACRO{\unit{m}}%
%BeginExpansion
\operatorname{m}%
%EndExpansion
$ less than the experimental data, especially in the temperature range $300%
%TCIMACRO{\unit{K}}%
%BeginExpansion
\operatorname{K}%
%EndExpansion
<T<425%
%TCIMACRO{\unit{K}}%
%BeginExpansion
\operatorname{K}%
%EndExpansion
$.%

%TCIMACRO{\TeXButton{Table6}{\begin{table}[b]
%\caption{Surface tension $\gamma_p$ and $\gamma
%$ for the SPC/E model as a function of LJ cutoff}
%\label{table6}
%\begin{ruledtabular}
%\begin{tabular}{ccccc}
%& \multicolumn{2}{c}{$\gamma_p$ (mN/m)}
%& \multicolumn{2}{c}{$\gamma$ (mN/m)} \\
%\cline{2-5}
%$T$ (K) &
%$r_c = 10$ \AA &
%$r_c = 16$ \AA &
%$r_c = 10$ \AA &
%$r_c = 16$ \AA \\
%\hline300&	49.9&	50.2&	55.4&	52.3\\
%325&	42.8&	48.7&	47.8&	50.7\\
%350&	42.2&	47.2&	47.0&	49.1\\
%375&	40.0&	40.3&	44.6&	42.1\\
%400&        33.3&	37.7&	37.6&       39.4\\
%425&        28.1&       33.5&       32.0&	35.0\\
%450&	27.0&	29.2&	30.6&	30.6\\
%475&	23.6&	23.6&	26.8&	24.9\\
%500&	20.5&	22.3&	23.2&	23.3
%\end{tabular}
%\end{ruledtabular}
%\end{table}}}%
%BeginExpansion
\begin{table}[b]
\caption{Surface tension $\gamma_p$ and $\gamma
$ for the SPC/E model as a function of LJ cutoff}
\label{table6}
\begin{ruledtabular}
\begin{tabular}{ccccc}
& \multicolumn{2}{c}{$\gamma_p$ (mN/m)}
& \multicolumn{2}{c}{$\gamma$ (mN/m)} \\
\cline{2-5}
$T$ (K) &
$r_c = 10$ \AA &
$r_c = 16$ \AA &
$r_c = 10$ \AA &
$r_c = 16$ \AA \\
\hline300&	49.9&	50.2&	55.4&	52.3\\
325&	42.8&	48.7&	47.8&	50.7\\
350&	42.2&	47.2&	47.0&	49.1\\
375&	40.0&	40.3&	44.6&	42.1\\
400&        33.3&	37.7&	37.6&       39.4\\
425&        28.1&       33.5&       32.0&	35.0\\
450&	27.0&	29.2&	30.6&	30.6\\
475&	23.6&	23.6&	26.8&	24.9\\
500&	20.5&	22.3&	23.2&	23.3
\end{tabular}
\end{ruledtabular}
\end{table}%
%EndExpansion
%

%TCIMACRO{\FRAME{ftbFU}{3.4255in}{4.8837in}{0pt}{\Qcb{(Top figure) Surface
%tension of the three-point models of water as a function of temperature: SPC/E
%(filled circles), TIP3P (squares), TIP3P-C (triangles), and TIP3P-Ew
%(diamonds). Simulation data from Alejandre \QTR{em}{et al.} \cite{Alejandre95}
%(open circles) are included for comparison. (Bottom figure) Surface tension of
%the four-point models of water as a function of temperature: TIP4P (circles)
%and TIP4P-Ew (squares). In both figures, experimental data
%\cite{Gittens69,Cini72,Jasper72} (solid curve) and extrapolation of quadratic
%fit to higher temperatures (dashed curve) are included for comparison.}}%
%{\Qlb{fig3ab}}{surfacetension.eps}{\special{ language "Scientific Word";
%type "GRAPHIC";  maintain-aspect-ratio TRUE;  display "USEDEF";
%valid_file "F";  width 3.4255in;  height 4.8837in;  depth 0pt;
%original-width 7.5296in;  original-height 9.4472in;  cropleft "0";
%filename 'SurfaceTension.eps';file-properties "XNPEU";}}}%
%BeginExpansion
%EndExpansion

Alejandre \emph{et al}. \cite{Alejandre95} and Shi \emph{et al. }\cite{Shi06}
reported excellent agreement with experimental results for the SPC/E model.
However, our results for the SPC/E model clearly disagree with their data as
well as with experimental results, although the simulations were performed
under essentially identical conditions with respect to the number of molecules
and the dimensions of the system, potentials employed, temperature range, and
cutoffs. We study the potential causes of the disagreement in the results
below in Sections \ref{cutoff} and \ref{rsad}.

Results for the four-point models as a function of temperature are shown in
Figure \ref{fig3ab}(b). The uncertainty for the four-point models is the same
as for the three-point models. Like the three-point models, the four-point
models underestimate the surface tension, with the TIP4P\ model offering
results comparable to the TIP3P-C and TIP3P-Ew models, while the performance
of the TIP4P-Ew model is significantly closer to the experimental data than
any of the other models examined here. Unlike the TIP3P models, at most
temperatures considered here, the TIP4P and TIP4P-Ew models do not agree
within simulation uncertainty.

\subsection{Tail correction}

While tail corrections can exist for both the Lennard-Jones and the
electrostatic interactions, by using Ewald summations, we avoid the need for a
Coulombic tail correction. In evaluating the tail correction, Eq.
(\ref{eqn15}) for the Lennard-Jones potential in the region $r\gg\sigma$,
$dU\left(  r\right)  /dr\approx24\varepsilon\sigma^{6}r^{-7}$. Assuming that
for $r>r_{c}$, the radial distribution function $g\left(  r\right)  \approx1$
and that the density profile is an error function of the form Eq.
(\ref{eqn20}), we can evaluate Eq. (\ref{eqn15}). After integration over $s$,%
\begin{multline}
\frac{\gamma_{tail}}{12\pi\varepsilon\sigma^{6}}=\int_{-\infty}^{\infty}%
\int_{r_{c}}^{\infty}\frac{\operatorname{erf}\left(  \xi z\right)  }{\sqrt
{\pi}\xi^{3}r^{7}}e^{-\xi^{2}\left(  r+z\right)  ^{2}}\times\label{eqn22}\\
\left(  \xi^{2}z\left(  r-z\right)  +e^{4\xi^{2}rz}\left(  \xi^{2}z\left(
z+r\right)  +1\right)  -1\right)  ~\\
-\frac{z\operatorname{erf}\left(  \xi z\right)  }{2\xi^{2}r^{7}}\left(
3-\xi^{2}\left(  z^{2}-r^{2}\right)  \right)  f\left(  \xi,r,z\right)  ~dr~dz,
\end{multline}
where $\xi\equiv\sqrt{\pi}/w_{e}$ and
\[
f\left(  \xi,r,z\right)  =\operatorname{erf}\left(  \xi\left(  r+z\right)
\right)  +\operatorname{erf}\left(  \xi\left(  r-z\right)  \right)  .
\]
Eq. (\ref{eqn22}) is then evaluated numerically for each density profile. For
the SPC/E model of water, the interfacial thickness, density difference, and
tail correction are shown in Table \ref{table3}. Because of the existence of
capillary waves at the interface, as shown in Section \ref{sec-cw}, the
interfacial thickness parameter $w_{e}$ depends logarithmically on the length
$L_{\parallel}$ of the interfacial surface$.$

The tail correction depends strongly upon the Lennard-Jones parameters,
scaling as $\varepsilon\sigma^{6}$, and decays exponentially as a function of
the chosen interaction cutoff $r_{c}$. Because $\varepsilon_{OO}$ and
$\sigma_{OO}$ are approximately equal for the SPC/E and TIP3P models, the tail
corrections at all temperatures are almost identical for the two models. While
the TIP3P-C has additional tail corrections for the O-H and H-H interactions,
their magnitudes are negligible in comparison to the correction for the O-O
interaction. Only the TIP3P-Ew model, which has a significantly smaller value
for the Lennard-Jones interaction strength $\varepsilon_{OO}$, has a
noticeably different tail correction from the other three-point models. The
tail corrections for the four-point models are likewise close to that of the
SPC/E model, with the TIP4P model having a slightly smaller tail correction
and the TIP4P-Ew model a slightly larger tail correction.

\subsection{\label{cutoff}Cutoff effects}

The original parameterizations for the SPC/E and TIP3P models of water
employed cutoffs for both electrostatic and Lennard-Jones interactions
\cite{Berendsen87,Jorgensen83}. To study the effect of varying the
electrostatic cutoff, we applied a short-range cutoff to both the LJ and
electrostatic potentials of the SPC/E model, truncating the potentials at
$r_{cut}=10$, $12 $, $14$, $16$, $18$, and $20%
%TCIMACRO{\unit{\mathring{A}}}%
%BeginExpansion
\operatorname{\mathring{A}}%
%EndExpansion
$. Using Eq. (\ref{eqn14}) to determine the surface tension, we found that the
estimated values of the surface tension were nonsensical, ranging between
$-3700$ and $1700%
%TCIMACRO{\unit{mN}}%
%BeginExpansion
\operatorname{mN}%
%EndExpansion%
%TCIMACRO{\unit{m}}%
%BeginExpansion
\operatorname{m}%
%EndExpansion
^{-1}$, with no value smaller in magnitude than $140%
%TCIMACRO{\unit{mN}}%
%BeginExpansion
\operatorname{mN}%
%EndExpansion%
%TCIMACRO{\unit{m}}%
%BeginExpansion
\operatorname{m}%
%EndExpansion
^{-1}$. This shows that truncated electrostatic potentials are inappropriate
for use in the determination of the surface tension of water.

To determine the effect of varying only the range of the Lennard-Jones
interaction on $\gamma_{p}$ before incorporating the tail correction, we
performed runs for the SPC/E model at $300%
%TCIMACRO{\unit{K}}%
%BeginExpansion
\operatorname{K}%
%EndExpansion
$ with LJ cutoffs of $10$, $12$, $14$, $16$, $18$, and $20%
%TCIMACRO{\unit{\mathring{A}}}%
%BeginExpansion
\operatorname{\mathring{A}}%
%EndExpansion
,$ using the PPPM\ Ewald technique for the electrostatic forces. The starting
configuration for these runs was the final configuration from the $300%
%TCIMACRO{\unit{K}}%
%BeginExpansion
\operatorname{K}%
%EndExpansion
$ simulation used to compute the surface tension in Section \ref{Tdep}. The
resulting values of the surface tension are shown in Table \ref{table4}. The
values for the surface tension are within the simulation uncertainties,
although they tend to rise with increasing $r_{c}$. This is further reflected
in the density profiles, which show that the liquid-phase density $\rho_{L}$
increases with $r_{c}$ for values of $r_{c}$ between $10%
%TCIMACRO{\unit{\mathring{A}}}%
%BeginExpansion
\operatorname{\mathring{A}}%
%EndExpansion
$ and $18%
%TCIMACRO{\unit{\mathring{A}}}%
%BeginExpansion
\operatorname{\mathring{A}}%
%EndExpansion
$. The overall effect of the choice of $r_{c}$ can be seen when comparing the
surface tensions of SPC/E water in the temperature range $300%
%TCIMACRO{\unit{K}}%
%BeginExpansion
\operatorname{K}%
%EndExpansion
$ to $500%
%TCIMACRO{\unit{K}}%
%BeginExpansion
\operatorname{K}%
%EndExpansion
$ for LJ cutoffs of $10%
%TCIMACRO{\unit{\mathring{A}}}%
%BeginExpansion
\operatorname{\mathring{A}}%
%EndExpansion
$ and $16%
%TCIMACRO{\unit{\mathring{A}}}%
%BeginExpansion
\operatorname{\mathring{A}}%
%EndExpansion
$. The resulting data are shown in Table \ref{table6}. The data demonstrate
that the surface tensions for $10%
%TCIMACRO{\unit{\mathring{A}}}%
%BeginExpansion
\operatorname{\mathring{A}}%
%EndExpansion
$ and $16%
%TCIMACRO{\unit{\mathring{A}}}%
%BeginExpansion
\operatorname{\mathring{A}}%
%EndExpansion
$ cutoffs are equal, within simulation uncertainty, after the corresponding
tail corrections have been applied to each set of data. Thus, the $10%
%TCIMACRO{\unit{\mathring{A}}}%
%BeginExpansion
\operatorname{\mathring{A}}%
%EndExpansion
$ Lennard-Jones cutoff with long-range tail corrections is sufficiently
accurate for computing the surface tension.

\subsection{\label{rsad}Reciprocal space accuracy dependence}

\begin{figure}
[tb]
\begin{center}
\includegraphics[
height=2.8791in,
width=3.4255in
]%
{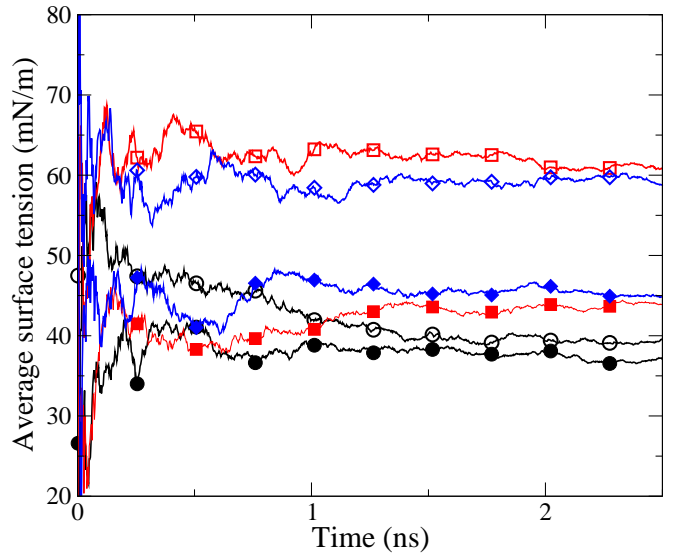}%
\caption{(Color online) Equilibration of the surface tension of the SPC/E
(black circles), TIP3P (red squares), and TIP3P-Ew (blue diamonds) models of
water for rms $\mathbf{k}$-space accuracies of $4\times10^{-3}$ (dashed
curves, open symbols) and $10^{-4}$ (solid curves, solid symbols).}%
\label{fig2}%
\end{center}
\end{figure}

Alejandre \emph{et al.} \cite{Alejandre95} found that the surface tension
depended on the mesh refinement $\left\vert \mathbf{h}^{\max}\right\vert $
used to evaluate the long-range Coulombic interactions. To test the dependence
of $\gamma_{p}$ on $\left\vert \mathbf{h}^{\max}\right\vert ,$ we show results
in Figure \ref{fig2} for three different models as a function of $\left\vert
\mathbf{h}^{\max}\right\vert $. From the figure, several trends become
apparent. First, for both the TIP3P and TIP3P-Ew models, the long-time average
of the surface tension depends significantly on the value $\left\vert
\mathbf{h}^{\max}\right\vert $:\ the long-time average for a $5\times
5\times20$ mesh ($\left\vert \mathbf{h}^{\max}\right\vert =20$, rms accuracy
$4.0\times10^{-3}$) is between $15$ and $20%
%TCIMACRO{\unit{mN}}%
%BeginExpansion
\operatorname{mN}%
%EndExpansion%
%TCIMACRO{\unit{m}}%
%BeginExpansion
\operatorname{m}%
%EndExpansion
^{-1}$ larger than for a $12\times12\times48$ mesh ($\left\vert \mathbf{h}%
^{\max}\right\vert =50$, rms accuracy $10^{-4}$). For larger values of
$\left\vert \mathbf{h}^{\max}\right\vert ,$ there is no significant adjustment
in the surface tension. Additionally, as found above, for a given value of the
precision, there is little difference in the long-time average of the two models.%

%TCIMACRO{\FRAME{ftbFU}{3.4255in}{2.8791in}{0pt}{\Qcb{(Color online)
%Equilibration of the surface tension of the SPC/E (black circles), TIP3P (red
%squares), and TIP3P-Ew (blue diamonds) models of water for rms $\QTR{bf}{k}$%
%-space accuracies of $4\times10^{-3}$ (dashed curves, open symbols) and
%$10^{-4}$ (solid curves, solid symbols).}}{\Qlb{fig2}}{kspace-accuracy.eps}%
%width 3.4255in;  height 2.8791in;  depth 0pt;  original-width 9.7577in;
%original-height 6.8581in;  cropleft "0";  croptop "1";  cropright "1";
%cropbottom "0";  filename 'kspace-accuracy.eps';file-properties "XNPEU";}}}%
%BeginExpansion
%EndExpansion

For the SPC/E model, there is little difference between the equilibrium values
for the $5\times5\times20$ mesh and the $12\times12\times48$ mesh. However, it
takes approximately $1%
%TCIMACRO{\unit{ns}}%
%BeginExpansion
\operatorname{ns}%
%EndExpansion
$ to achieve agreement between the two precision levels; before this, the
less-refined mesh has a significantly greater surface tension. In the work of
Alejandre \emph{et al}., the total simulation time was only $0.3\allowbreak75%
%TCIMACRO{\unit{ns}}%
%BeginExpansion
\operatorname{ns}%
%EndExpansion
.$ After $0.375%
%TCIMACRO{\unit{ns}}%
%BeginExpansion
\operatorname{ns}%
%EndExpansion
$, the average surface tension from our simulation using the $5\times
5\times20$ mesh was approximately $\gamma=60%
%TCIMACRO{\unit{mN}}%
%BeginExpansion
\operatorname{mN}%
%EndExpansion%
%TCIMACRO{\unit{m}}%
%BeginExpansion
\operatorname{m}%
%EndExpansion
^{-1}$, which closely corresponds with the result $\gamma=60.6%
%TCIMACRO{\unit{mN}}%
%BeginExpansion
\operatorname{mN}%
%EndExpansion%
%TCIMACRO{\unit{m}}%
%BeginExpansion
\operatorname{m}%
%EndExpansion
^{-1}$ obtained in the earlier study. However, for long simulation times, the
surface tension of the SPC/E model does not exhibit a strong dependence on the
mesh size. Consequently, since the long-time averages are essentially equal
within simulation uncertainty, we have used the finer $12\times12\times48$
mesh refinement for all of the simulations reported in this paper unless
otherwise specified.%

%TCIMACRO{\FRAME{ftbFU}{3.4255in}{2.4561in}{0pt}{\Qcb{Surface tension for the
%SPC/E model of water as a function of mesh accuracy averaged over (top)
%125-375 ps and (bottom) 1-2 ns for a $5\times5\times20$ mesh (circles), a
%$12\times12\times48$ mesh (diamonds), and a $20\times20\times80$ mesh
%(triangles).}}{\Qlb{fig6}}{accuracy.eps}%
%width 3.4255in;  height 2.4561in;  depth 0pt;  original-width 9.6033in;
%original-height 6.8581in;  cropleft "0";  croptop "1";  cropright "1";
%cropbottom "0";  filename 'accuracy.eps';file-properties "XNPEU";}}}%
%BeginExpansion
\begin{figure}
[tb]
\begin{center}
\includegraphics[
height=2.4561in,
width=3.4255in
]%
{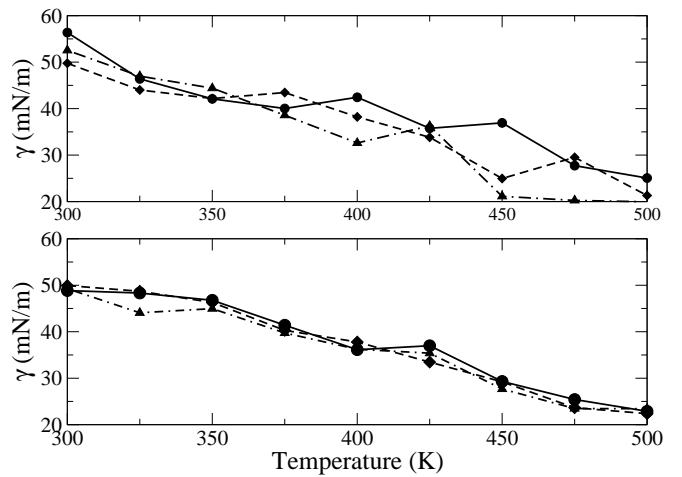}%
\caption{Surface tension for the SPC/E model of water as a function of mesh
accuracy averaged over (top) 125-375 ps and (bottom) 1-2 ns for a
$5\times5\times20$ mesh (circles), a $12\times12\times48$ mesh (diamonds), and
a $20\times20\times80$ mesh (triangles).}%
\label{fig6}%
\end{center}
\end{figure}
%EndExpansion

To further demonstrate that the accuracy of the reciprocal-space calculation
does not affect the results for the surface tension of the SPC/E model when
averaged over sufficiently long times, we performed simulations of the SPC/E
model, as described above, with $\mathbf{k}$-space meshes of $5\times
5\times20$, $12\times12\times48$, and $20\times20\times80,$ corresponding to
rms accuracies of $4\times10^{-3}$, $10^{-4}$, and $10^{-5}$, respectively.
The results are shown in Figure \ref{fig6} for the time ranges of $125$ to
$375%
%TCIMACRO{\unit{ps}}%
%BeginExpansion
\operatorname{ps}%
%EndExpansion
,$ used by Alejandre \emph{et al.}, and $1$ to $2%
%TCIMACRO{\unit{ns}}%
%BeginExpansion
\operatorname{ns}%
%EndExpansion
,$ used in the present work. Clearly, the results for the shorter time range
($125$ to $375%
%TCIMACRO{\unit{ps}}%
%BeginExpansion
\operatorname{ps}%
%EndExpansion
$) have not converged: there is significant disagreement of as much as $20%
%TCIMACRO{\unit{mN}}%
%BeginExpansion
\operatorname{mN}%
%EndExpansion
/%
%TCIMACRO{\unit{m}}%
%BeginExpansion
\operatorname{m}%
%EndExpansion
$, particularly at higher temperatures. However, for longer times, the results
have converged, with the differences among the three mesh refinements being
essentially within simulation uncertainty. It is interesting to note that the
converged surface tensions of the different mesh refinements corresponds very
closely to the profile obtained for the intermediate mesh refinement
($12\times12\times48$) after $375%
%TCIMACRO{\unit{ps}}%
%BeginExpansion
\operatorname{ps}%
%EndExpansion
$. Our intermediate mesh refinement is comparable to the most refined
$\mathbf{k}$-space mesh considered by Alejandre \emph{et al. }%
\cite{Alejandre95} (shown in Fig. 7 of their paper for a single, instantaneous configuration).%

%TCIMACRO{\TeXButton{Table5}{\begin{table}[tb]
%\caption
%{Comparison of surface tension for the SPC/E model as a function of calculation method}
%\label{table5}
%\begin{tabular}{@{\extracolsep{0.25in}}cccc}
%\hline\hline\
%& \multicolumn{3}{c}{Surface tension, mN/m} \\
%\cline{2-4}
%& \multicolumn{2}{c}{Capillary-wave} \\
%\cline{2-3}
%$T$ (K) & Erf, $\gamma_w$ & Tanh, $\gamma_w$
%& Pressure-integration, $\gamma_p$  \\
%\hline300 & 46.1 & 36.5 & 49.9\\
%400 & 32.0 & 27.8 & 33.1 \\
%500 & 19.0 & 19.0 & 20.2 \\
%\hline\hline\
%\end{tabular}
%\end{table}}}%
%BeginExpansion
\begin{table}[tb]
\caption
{Comparison of surface tension for the SPC/E model as a function of calculation method}
\label{table5}
\begin{tabular}{@{\extracolsep{0.25in}}cccc}
\hline\hline\
& \multicolumn{3}{c}{Surface tension, mN/m} \\
\cline{2-4}
& \multicolumn{2}{c}{Capillary-wave} \\
\cline{2-3}
$T$ (K) & Erf, $\gamma_w$ & Tanh, $\gamma_w$
& Pressure-integration, $\gamma_p$  \\
\hline300 & 46.1 & 36.5 & 49.9\\
400 & 32.0 & 27.8 & 33.1 \\
500 & 19.0 & 19.0 & 20.2 \\
\hline\hline\
\end{tabular}
\end{table}%
%EndExpansion
%

%TCIMACRO{\TeXButton{Table7}{\begin{table}[tb]
%\caption
%{Surface tension for the SPC/E model as a function of system size at 300 K}
%\label{table7}
%\begin{tabular}{@{\extracolsep{0.25in}}ccc}
%\hline\hline\
%Interfacial length $L_{\parallel}$, \AA& $\gamma_p$ (mN/m)
%& $\gamma$ (mN/m) \\
%\hline2.3 & 49.9 & 55.4 \\
%9.2 & 52.0 & 57.5 \\
%11.5 & 51.5 & 57.0\\
%13.8 & 51.8 & 57.3\\
%16.1 & 51.9 & 57.4 \\
%23.0 & 51.6 & 57.1 \\
%34.5 & 52.0 & 57.5 \\
%46.0 & 51.8 & 57.3 \\
%\hline\hline\
%\end{tabular}
%\end{table}}}%
%BeginExpansion
\begin{table}[tb]
\caption
{Surface tension for the SPC/E model as a function of system size at 300 K}
\label{table7}
\begin{tabular}{@{\extracolsep{0.25in}}ccc}
\hline\hline\
Interfacial length $L_{\parallel}$, \AA& $\gamma_p$ (mN/m)
& $\gamma$ (mN/m) \\
\hline2.3 & 49.9 & 55.4 \\
9.2 & 52.0 & 57.5 \\
11.5 & 51.5 & 57.0\\
13.8 & 51.8 & 57.3\\
16.1 & 51.9 & 57.4 \\
23.0 & 51.6 & 57.1 \\
34.5 & 52.0 & 57.5 \\
46.0 & 51.8 & 57.3 \\
\hline\hline\
\end{tabular}
\end{table}%
%EndExpansion

\section{\label{sec-cw}Capillary waves}

\begin{figure}
[tb]
\begin{center}
\includegraphics[
height=2.4037in,
width=3.4255in
]%
{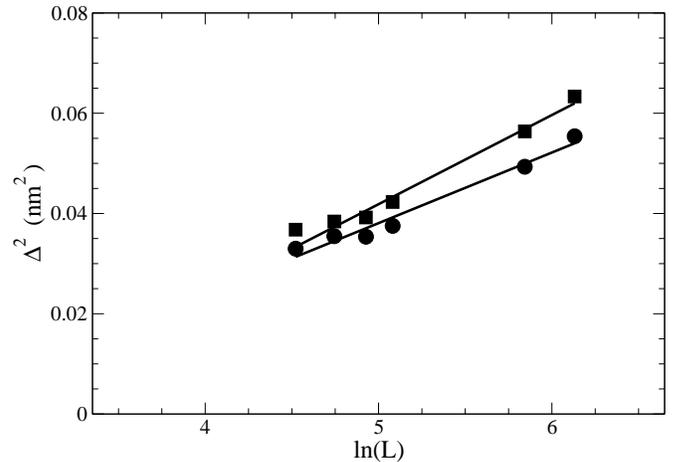}%
\caption{Regression fit of Eq. (\ref{eqn17}) for hyperbolic tangent (diamonds)
and error function (circles) profiles at $300\operatorname{K}$.}%
\label{fig4}%
\end{center}
\end{figure}

The interfacial width $\Delta^{2}$ was computed for various system sizes as
described in Section \ref{mm}. The resulting plots of $\Delta^{2}$ versus $\ln
L$ were computed and the value of $\gamma_{w}$ extracted using Eq.
(\ref{eqn17}). A plot showing the data obtained for $300%
%TCIMACRO{\unit{K}}%
%BeginExpansion
\operatorname{K}%
%EndExpansion
$ for the SPC/E water model is shown in Figure \ref{fig4}; the resulting
values of $\gamma_{w}$ for the two functional forms at $300%
%TCIMACRO{\unit{K}}%
%BeginExpansion
\operatorname{K}%
%EndExpansion
$, $400%
%TCIMACRO{\unit{K}}%
%BeginExpansion
\operatorname{K}%
%EndExpansion
$, and $500%
%TCIMACRO{\unit{K}}%
%BeginExpansion
\operatorname{K}%
%EndExpansion
$ are shown in Table \ref{table5}.%

%TCIMACRO{\FRAME{ftbFU}{3.4255in}{2.4037in}{0pt}{\Qcb{Regression fit of Eq.
%(\ref{eqn17}) for hyperbolic tangent (diamonds) and error function (circles)
%profiles at $300\unit{K}$.}}{\Qlb{fig4}}{fits4.eps}%
%width 3.4255in;  height 2.4037in;  depth 0pt;  original-width 9.9608in;
%original-height 6.8732in;  cropleft "0";  croptop "1";  cropright "1";
%cropbottom "0";  filename 'fits4.eps';file-properties "XNPEU";}}}%
%BeginExpansion
%EndExpansion

Our results indicate that at all three temperatures, good agreement between
the pressure-integration estimate of the surface tension, $\gamma_{p}$, and
the capillary-wave estimate of the surface tension, $\gamma_{w}$, is obtained
only if the interfacial density profile is fit to an error function. There is
substantial disagreement between the error function and hyperbolic tangent
functions at lower temperatures: the hyperbolic tangent profile always yields
lower estimates than the error function profile. However, the magnitude of the
discrepancy between the two estimates of $\gamma$ decreases the two narrows as
temperature increases, and essentially vanishes at $500%
%TCIMACRO{\unit{K}}%
%BeginExpansion
\operatorname{K}%
%EndExpansion
$.

In addition to computing the surface tension $\gamma_{w}$, we can also compute
an upper-bound estimate for the intrinsic interfacial width $\Delta_{0}$
\cite{Lacasse98}. After obtaining the slope $\alpha_{w}$ and intercept
$\beta_{w}$ for the least-squares fit of Eq. (\ref{eqn17}), we assume that the
parameter $B_{0}=c\Delta_{0}$, where $c$ is a constant to be specified. The
intercept $\beta_{w}$ and $\Delta_{0}$ are then related by%
\begin{equation}
\beta_{w}=\Delta_{0}^{2}-\alpha_{w}\ln\left(  c\Delta_{0}\right)  .
\label{eqn24}%
\end{equation}
For values of $c$ less than some threshold $c^{\ast}$, there is no real
solution to (\ref{eqn24}); above the threshold, $\Delta_{0}$ quickly decays as
$c$ increases. Thus, $\Delta_{0}$ has a maximum at the threshold value
$c=c^{\ast}$ where the imaginary part of the solution vanishes. For the system
sizes under consideration, we find that the maximum intrinsic width
$\Delta_{0}\approx0.8%
%TCIMACRO{\unit{\mathring{A}}}%
%BeginExpansion
\operatorname{\mathring{A}}%
%EndExpansion
,$ $1.0%
%TCIMACRO{\unit{\mathring{A}}}%
%BeginExpansion
\operatorname{\mathring{A}}%
%EndExpansion
$, and $1.5%
%TCIMACRO{\unit{\mathring{A}}}%
%BeginExpansion
\operatorname{\mathring{A}}%
%EndExpansion
$ at $T=300%
%TCIMACRO{\unit{K}}%
%BeginExpansion
\operatorname{K}%
%EndExpansion
$, $400%
%TCIMACRO{\unit{K}}%
%BeginExpansion
\operatorname{K}%
%EndExpansion
$, and $500%
%TCIMACRO{\unit{K}}%
%BeginExpansion
\operatorname{K}%
%EndExpansion
$.

\begin{figure}
[tb]
\begin{center}
\includegraphics[
height=2.4082in,
width=3.4255in
]%
{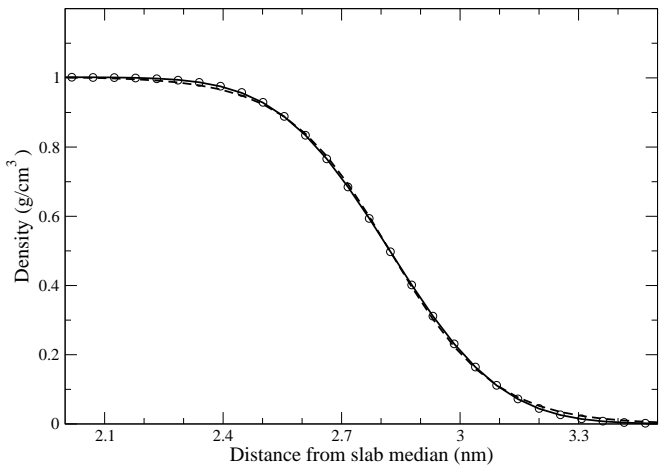}%
\caption{Density profile in the interfacial region for a slab of 400,000 SPC/E
water molecules at $300\operatorname{K}$ ($L_{\parallel}=46.0\operatorname{nm}%
$). Simulation results are shown as circles; fits to error function (solid
curve)\ and hyperbolic tangent function (dashed curve)\ are also included.}%
\label{fig8}%
\end{center}
\end{figure}

\begin{figure}
[tb]
\begin{center}
\includegraphics[
height=2.3044in,
width=3.4246in
]%
{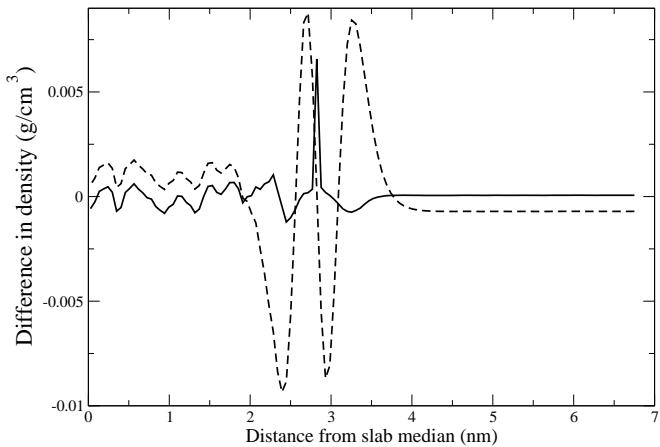}%
\caption{Difference between simulation results and the error function (solid
curve) and the hyperbolic tangent (dashed curve) fits for the same system
shown in Figure \ref{fig8}.}%
\label{fig9}%
\end{center}
\end{figure}

Further evidence that fitting the density profile to an error function yields
more accurate results than fitting to a hyperbolic tangent function can be
seen by comparing the density fits themselves. As shown in Figure \ref{fig8},
although the two profiles are similar, the error function fit more closely
adheres to simulation results than the hyperbolic tangent profile. Although
the $\chi^{2}$ parameter for both functional fits was relatively small, the
coefficient for the hyperbolic tangent $\chi^{2}$ parameter ($\sim10^{-3}$)
was approximately two orders of magnitude larger than the corresponding error
function $\chi^{2}$ parameter ($\sim10^{-5}$). The greater accuracy of the
error function is further seen by comparing the magnitude of the differences
between the simulation results and the fitted functional profiles, as shown in
Figure \ref{fig9}.

In addition to observing capillary wave behavior, the larger simulations can
be used to study the effect of the interfacial area $L_{\parallel}^{2}$ on the
surface tension computed using Eqs. (\ref{eqn14}) and (\ref{eqn22}). Since
surface tension is an intensive property, it should be independent of
$L_{\parallel}^{2}$. Although Eq. (\ref{eqn22}) is a slowly decreasing
function of the parameter $w_{e}$, over the range of values for $w_{e}$ and
$\Delta^{2}$ considered here, the tail contribution $\gamma_{tail}$ varies by
less than 2\%. Thus, we expect the simulation values for the surface tension
to remain constant, independent of $L_{\parallel}$. Examining the results for
the surface tension $\gamma_{p}$ (without tail correction) versus the system
size $N$, as shown in Table \ref{table7}, we find a slight increase in
$\gamma$ as $L_{\parallel}$ increases, although the results remain within
simulation uncertainty, even for large values of $L_{\parallel}$.%

%TCIMACRO{\FRAME{ftbFU}{3.4255in}{2.4082in}{0pt}{\Qcb{Density profile in the
%interfacial region for a slab of 400,000 SPC/E water molecules at
%$300\unit{K}$ ($L_{\parallel}=46.0\unit{nm}$). Simulation results are shown as
%circles; fits to error function (solid curve)\ and hyperbolic tangent function
%(dashed curve)\ are also included.}}{\Qlb{fig8}}{300kcw_comp.eps}%
%width 3.4255in;  height 2.4082in;  depth 0pt;  original-width 9.6317in;
%original-height 6.7446in;  cropleft "0";  croptop "1";  cropright "1";
%cropbottom "0";  filename '300Kcw_comp.eps';file-properties "XNPEU";}}}%
%BeginExpansion
%EndExpansion
%

%TCIMACRO{\FRAME{ftbFU}{3.4246in}{2.3044in}{0pt}{\Qcb{Difference between
%simulation results and the error function (solid curve) and the hyperbolic
%tangent (dashed curve) fits for the same system shown in Figure \ref{fig8}.}%
%}{\Qlb{fig9}}{300kcw_diff.eps}{\special{ language "Scientific Word";
%type "GRAPHIC";  maintain-aspect-ratio TRUE;  display "USEDEF";
%valid_file "F";  width 3.4246in;  height 2.3044in;  depth 0pt;
%original-width 9.9298in;  original-height 6.7446in;  cropleft "0";
%filename '300Kcw_diff.eps';file-properties "XNPEU";}}}%
%BeginExpansion
%EndExpansion

\section{Conclusions}

An investigation of surface tension as a function of temperature for a number
of popular three- and four-point water models shows systematic disagreement
between experimental and simulation results for all six models considered. All
six models considered consistently underestimate the surface tension relative
to the experimental data. The TIP4P-Ew model is closest to the experimental
data, although not in quantitative agreement.

Examining the SPC/E model in greater detail, we have illustrated the
importance of having a sufficiently long simulation time and sufficiently fine
$\mathbf{k}$-space mesh: significant variations in the surface tension can
result if the equilibration period is too short. In the SPC/E, TIP3P, and
TIP3P-Ew models, significant variations were also observed if a small number
of $\mathbf{k}$-space vectors are used.

The present study also includes the first in-depth study of the effect of
large system sizes on interfacial properties, studying capillary waves for
systems of up to $4\times10^{5}$ molecules. We have demonstrated that
examining capillary waves at the liquid-vapor interface can be used to
determine the surface tension of real fluids, and that the use of an error
function profile offers better results in comparison to experimental data than
the use of a hyperbolic tangent profile. Finally, we note that the interfacial
width $\Delta$ depends on the interfacial length $L_{\parallel}$, and that
attempting to extract an interfacial width without taking into account the
effect of capillary waves \cite{Kuo06} is incorrect.%

%TCIMACRO{\TeXButton{Acknowledgments}{\section*{Acknowledgments}}}%
%BeginExpansion
\section*{Acknowledgments}%
%EndExpansion

The authors would like to thank Steve Plimpton, Paul Crozier, and Amalie
Frischknecht for helpful discussions in implementing the TIP4P model. Sandia
is a multiprogram laboratory operated by Sandia Corporation, a Lockheed Martin
Company, for the United States Department of Energy under Contract No. DE-AC04-94AL85000.

\bibliographystyle{jcc}
\bibliography{,,,,WATER}

\begin{thebibliography}{10}

\bibitem{Buff65}
F.~P. Buff, R.~A. Lovett, and F.~H. Stillinger, {\em Phys. Rev. Lett.}, {\bf
  15}, 621--623 (1965).

\bibitem{Rowlinson82}
J.~Rowlinson and B.~Widom, {\em Molecular Theory of Capillarity}, Clarendon
  Press, Oxford, 1982.

\bibitem{Mecke99}
K.~R. Mecke and S.~Dietrich, {\em Phys. Rev. E}, {\bf 59}, 6766--6784 (1999).

\bibitem{Senapati01}
S.~Senapati and M.~L. Berkowitz, {\em Phys. Rev. Lett.}, {\bf 87}, 176101
  (2001).

\bibitem{Rivera03}
J.~L. Rivera, C.~McCabe, and P.~T. Cummings, {\em Phys. Rev. E}, {\bf 67},
  011603 (2003).

\bibitem{Nijmeijer88}
M.~Nijmeijer, A.~Bakker, C.~Bruin, and J.~Sikkenk, {\em J, Chem. Phys.}, {\bf
  89}, 3789 (1988).

\bibitem{Adams91}
P.~Adams and J.~Henderson, {\em Mol. Phys.}, {\bf 73}, 1383 (1991).

\bibitem{Lacasse98}
M.-D. Lacasse, G.~S. Grest, and A.~J. Levine, {\em Phys. Rev. Lett.}, {\bf 80},
  309--312 (1998).

\bibitem{Guillot02}
B.~Guillot, {\em J. Molec. Liq.}, {\bf 101}, 219--260 (2002).

\bibitem{Rivera02}
J.~L. Rivera, M.~Predota, A.~A. Chialvo, and P.~T. Cummings, {\em Chem. Phys.
  Lett.}, {\bf 357}, 189--194 (2002).

\bibitem{Price04}
D.~J. Price and C.~L. Brooks~III, {\em J. Chem. Phys.}, {\bf 121}, 10096--11003
  (2004).

\bibitem{Horn04}
H.~W. Horn, W.~C. Swope, J.~W. Pitera, J.~D. Madera, T.~J. Dick, G.~L. Hura,
  and T.~Head-Gordon, {\em J. Chem. Phys.}, {\bf 120}, 9665--9678 (2004).

\bibitem{Jorgensen05}
W.~L. Jorgensen and J.~Tirado-Rives, {\em Proc. Natl. Acad. Sci.}, {\bf 102},
  6665--6670 (2005).

\bibitem{Berendsen81}
H.~J.~C. Berendsen, J.~P.~M. Postma, W.~F. Van~Gunsteren, and J.~Hermans In
  {\em Intermolecular Forces}, B.~Pullman, Ed.;
\newblock Reidel, Dordrecht, 1981;
\newblock page 331.

\bibitem{Stern01}
H.~A. Stern, F.~Rittner, B.~J. Berne, and R.~A. Friesner, {\em J. Chem. Phys.},
  {\bf 115}, 2237--2251 (2001).

\bibitem{Berendsen87}
H.~J.~C. Berendsen, J.~R. Grigera, and T.~P. Straatsma, {\em J. Phys. Chem.},
  {\bf 91}, 6269--6271 (1987).

\bibitem{Jorgensen83}
W.~L. Jorgensen, J.~Chandrasekhar, J.~D. Madura, R.~W. Impey, and M.~L. Klein,
  {\em J. Chem. Phys.}, {\bf 79}, 926--935 (1983).

\bibitem{Watanabe89}
K.~Watanabe and M.~L. Klein, {\em Chem. Phys.}, {\bf 131}, 157--167 (1989).

\bibitem{Mahoney00}
M.~W. Mahoney and W.~L. Jorgensen, {\em J. Chem. Phys.}, {\bf 112}, 8910--8922
  (2000).

\bibitem{Bernal33}
J.~D. Bernal and R.~H. Fowler, {\em J. Chem. Phys.}, {\bf 1}, 515 (1933).

\bibitem{Stillinger74}
F.~H. Stillinger and A.~Rahman, {\em J. Chem. Phys.}, {\bf 60}, 1545--1557
  (1974).

\bibitem{Rick94}
S.~W. Rick, S.~J. Stuart, and B.~J. Berne, {\em J. Chem. Phys.}, {\bf 101},
  6141--6156 (1994).

\bibitem{Kuo06}
I.-F.~W. Kuo, C.~J. Mundy, B.~L. Eggimann, M.~J. McGrath, J.~I. Siepmann,
  B.~Chen, J.~Vieceli, and D.~J. Tobias, {\em J. Phys. Chem. B}, {\bf 110},
  3738--3746 (2006).

\bibitem{Mark01}
P.~Mark and L.~Nilsson, {\em J. Phys. Chem. A}, {\bf 105}, 9954--9960 (2001).

\bibitem{Mark02}
P.~Mark and L.~Nilsson, {\em J. Comp. Chem.}, {\bf 23}, 1211--1219 (2002).

\bibitem{Gittens69}
G.~J. Gittens, {\em J. Coll. Interf. Sci.}, {\bf 30}, 406--412 (1969).

\bibitem{Cini72}
R.~Cini, G.~Loglio, and A.~Ficaldi, {\em J. Coll. Interf. Sci.}, {\bf 41},
  287--298 (1972).

\bibitem{Jasper72}
J.~J. Jasper, {\em J. Phys. Chem. Ref. Data}, {\bf 1}, 841 (1972).

\bibitem{Johansson72}
K.~Johansson and J.~C. Eriksson, {\em J. Coll. Interf. Sci.}, {\bf 40},
  398--405 (1972).

\bibitem{Thakur75}
D.~K. Thakur and K.~Hickman, {\em J. Coll. Interf. Sci.}, {\bf 50}, 525--531
  (1975).

\bibitem{Alejandre95}
J.~Alejandre, D.~J. Tildesley, and G.~A. Chapela, {\em J. Chem. Phys.}, {\bf
  102}, 4574--4583 (1995).

\bibitem{Feller96}
S.~E. Feller, R.~W. Pastor, A.~Rojnuckarin, S.~Bogusz, and B.~R. Brooks, {\em
  J. Chem. Phys.}, {\bf 100}, 17011--17020 (1996).

\bibitem{Taylor96}
R.~S. Taylor, L.~X. Dang, and B.~C. Garrett, {\em J. Phys. Chem.}, {\bf 100},
  11720--11725 (1996).

\bibitem{Dang97}
L.~X. Dang and T.-M. Chang, {\em J. Chem. Phys.}, {\bf 106}, 8149--8159 (1997).

\bibitem{Zakharov97}
V.~V. Zakharov, E.~N. Brodskaya, and A.~Laaksonen, {\em J. Chem. Phys.}, {\bf
  107}, 10675--10683 (1997).

\bibitem{Zakharov98}
V.~V. Zakharov, E.~N. Brodskaya, and A.~Laaksonen, {\em Mol. Phys.}, {\bf 95},
  203--209 (1998).

\bibitem{Shi06}
B.~Shi, S.~Sinha, and V.~K. Dhir, {\em J. Chem. Phys.}, page In press.

\bibitem{Huang01}
D.~M. Huang, P.~L. Geissler, and D.~Chandler, {\em J. Phys. Chem. B}, {\bf
  105}, 6704--6709 (2001).

\bibitem{Tolman48}
R.~C. Tolman, {\em J. Chem. Phys.}, {\bf 16}, 758--774 (1948).

\bibitem{Kirkwood49}
J.~G. Kirkwood and F.~P. Buff, {\em J. Chem. Phys.}, {\bf 17}, 338--343 (1949).

\bibitem{Chapela77}
G.~A. Chapela, G.~Saville, S.~M. Thompson, and J.~S. Rowlinson, {\em J. Chem.
  Soc. Farad. Trans.}, {\bf 73}, 1133--1144 (1977).

\bibitem{Blokhuis95}
E.~M. Blokhuis, D.~Bedeaux, C.~D. Holcomb, and J.~A. Zollweg, {\em Mol. Phys.},
  {\bf 85}, 665--669 (1995).

\bibitem{Huang69}
J.~S. Huang and W.~W. Webb, {\em J. Chem. Phys.}, {\bf 50}, 3677--3693 (1969).

\bibitem{Beysens87}
D.~Beysens and M.~Robert, {\em J. Chem. Phys.}, {\bf 87}, 3056--3061 (1987).

\bibitem{Semenov94}
A.~N. Semenov, {\em Macromolecules}, {\bf 27}, 2732--2735 (1994).

\bibitem{Sides99}
S.~W. Sides, G.~S. Grest, and M.-D. Lacasse, {\em Phys. Rev. E}, {\bf 60},
  6708--6713 (1999).

\bibitem{Durell94}
S.~R. Durell, B.~R. Brooks, and A.~Ben-Naim, {\em J. Phys. Chem.}, {\bf 98},
  2198--2202 (1994).

\bibitem{Feenstra99}
K.~A. Feenstra, B.~Hess, and H.~J.~C. Berendsen, {\em J. Comp. Chem.}, {\bf
  20}, 786--798 (1999).

\bibitem{Spoel05}
D.~Van~der Spoel, E.~Lindahl, B.~Hess, A.~R. Buuren, E.~Apol, P.~J. Meulenhoff,
  D.~P. Tieleman, A.~L. T.~M. Sijbers, K.~A. Feenstra, R.~Van~Drunen, and
  H.~J.~C. Berendsen, {\em Gromacs User Manual Version 3.3}, www.gromacs.org,
  2005.

\bibitem{Plimpton95}
S.~J. Plimpton, {\em J. Comp. Phys.}, {\bf 117}, 1--9 (1995). See also 
  http://www.cs.sandia.gov/\~{}sjplimp/lammps.html

\bibitem{Ryckaert77}
J.-P. Ryckaert, G.~Ciccotti, and H.~J.~C. Berendsen, {\em J. Comp. Phys.}, {\bf
  23}, 327 (1977).

\bibitem{Hockney88}
R.~W. Hockney and J.~W. Eastwood, {\em Computer Simulation Using Particles},
  Adam Hilger-IOP Publishing, Bristol, 1988.

\bibitem{Flyvbjerg89}
H.~Flyvbjerg and H.~G. Petersen, {\em J. Chem. Phys.}, {\bf 91}, 461--466
  (1989).

\end{thebibliography}

\end{document}